\documentclass[conference]{IEEEconf}

\input epsf
\usepackage{graphicx}
\usepackage{multirow}
\usepackage{titlesec}
\usepackage{balance} 
\usepackage{stfloats}
\usepackage{xcolor}
\usepackage{url}

\renewcommand\thesection{\arabic{section}} 

\renewcommand\thesubsectiondis{\thesection.\arabic{subsection}}

\renewcommand\thesubsubsectiondis{\thesubsectiondis.\arabic{subsubsection}}

\begin{document}

\title{\textbf{\Large Multi-role Consensus through LLMs Discussions for Vulnerability Detection\\}}
%

\author{Zhenyu Mao$^{1}$, Jialong Li$^{1,*}$, 
Dongming Jin$^{2}$, 
Munan Li$^{3}$, and Kenji Tei$^{4}$\\
	\normalsize $^{1}$
 Waseda University, Tokyo, Japan
 $^{2}$
 Peking University, Beijing, China\\
 	\normalsize $^{3}$
  Dalian Maritime University, Dalian, China
  $^{4}$
 Tokyo Institute of Technology, Tokyo, Japan\\
	\normalsize *Corresponding Author: lijialong@fuji.waseda.jp
}


\maketitle
\begin{abstract}
Recent advancements in large language models (LLMs) have highlighted the potential for vulnerability detection, a crucial component of software quality assurance.
Despite this progress, most studies have been limited to the perspective of a single role, usually testers, lacking diverse viewpoints from different roles in a typical software development life-cycle, including both developers and testers.
To this end, this paper introduces a multi-role approach to employ LLMs to act as different roles simulating a real-life code review process and engaging in discussions toward a consensus on the existence and classification of vulnerabilities in the code.
Preliminary evaluation of this approach indicates a 13.48\% increase in the precision rate, an 18.25\% increase in the recall rate, and a 16.13\% increase in the F1 score.


\end{abstract}
\IEEEoverridecommandlockouts
\vspace{1.5ex}

%
\IEEEpeerreviewmaketitle

\section{Introduction}
As a fundamental process in software engineering, vulnerability detection plays a critical role in ensuring the quality and security of software systems by identifying potential security threats~\cite{li2024effectiveness_short,steenhoek2023dataflow_short}.
Nowadays, with the emergence of large language models (LLMs), due to their profound code comprehension, LLMs have also been employed for the application in the realm of vulnerability detection~\cite{zhang2023promptenhanced_short,nong2024chainofthought_short,zhou2024large_short}.

However, a notable limitation in these studies is that they often only consider the dominant tester-centric perspective in the prompt, such as asking LLMs to act as quality assurance engineers to detect potential vulnerabilities.
This single-role approach results in an imperfect understanding of the intentions behind the code, an incomplete exploration of potential issues, and consequently, reduced detection efficiency.
To avoid this, the real-world code review process usually involves the collaboration of team members (e.g., developer and tester) possessing diverse roles, responsibilities, and viewpoints.

Inspired by this, this paper introduces a novel multi-role approach, namely Multi-role Consensus through LLMs Discussions, for employing LLMs to act as multiple different roles in vulnerability detection, simulating a real-life code review process.
The key idea of this approach is to integrate the diverse perspectives of team members with different roles, thereby reaching a well-informed and collective consensus regarding the classification of the vulnerability in the code.

\section{Multi-role Approach}
\textbf{Initialization stage:}
The initiation stage is designed to enable the tester to independently provide an initial judgment.
The tester receives the initial prompt detailing its role-setting, its task, and the code segment to analyze.
The tester is asked to output its first response, a textual completion that includes a judgment, restricted to a binary indicator (1 for vulnerable, 0 for non-vulnerable), and a brief reasoning.
The response is constrained by a maximum token limit, ensuring that the tester's reasoning is both precise and substantive, facilitating clarity and efficiency in the subsequent dialectic interaction.
Then this initial judgment with reasoning is forwarded to the developer, together with a similar initial prompt.

\textbf{Discussion stage:}
The discussion stage aims to realize an iterative output exchange in an attempt to reach a collectively multi-perspective consensus inside the code review team.
The tester and the developer, equipped with their unique perspective and judgments, enter a dialectic interaction, aimed at exploring and resolving different opinions on potential vulnerabilities.
During this stage, the tester and the developer repeat a "pose query - deduce response - relay insight" loop, which serves as an incremental prompt, pushing participants to re-evaluate and polish their judgment and reasoning.
A maximum depth on discussion rounds is pre-set to prevent the dialogues from devolving into an endless cycle, ensuring that the discussion remains both goal-oriented and time-efficient.

\textbf{Conclusion stage:}
The conclusion stage summarizes the discussions and outputs the final result.
Once a consensus or the pre-set maximum discussion depth is reached, the tester's latest judgment is recorded as the final judgment, as the tester usually holds primary responsibility in the review process.

\begin{figure}[hbtp]
    \centering
    \includegraphics[height=28mm]{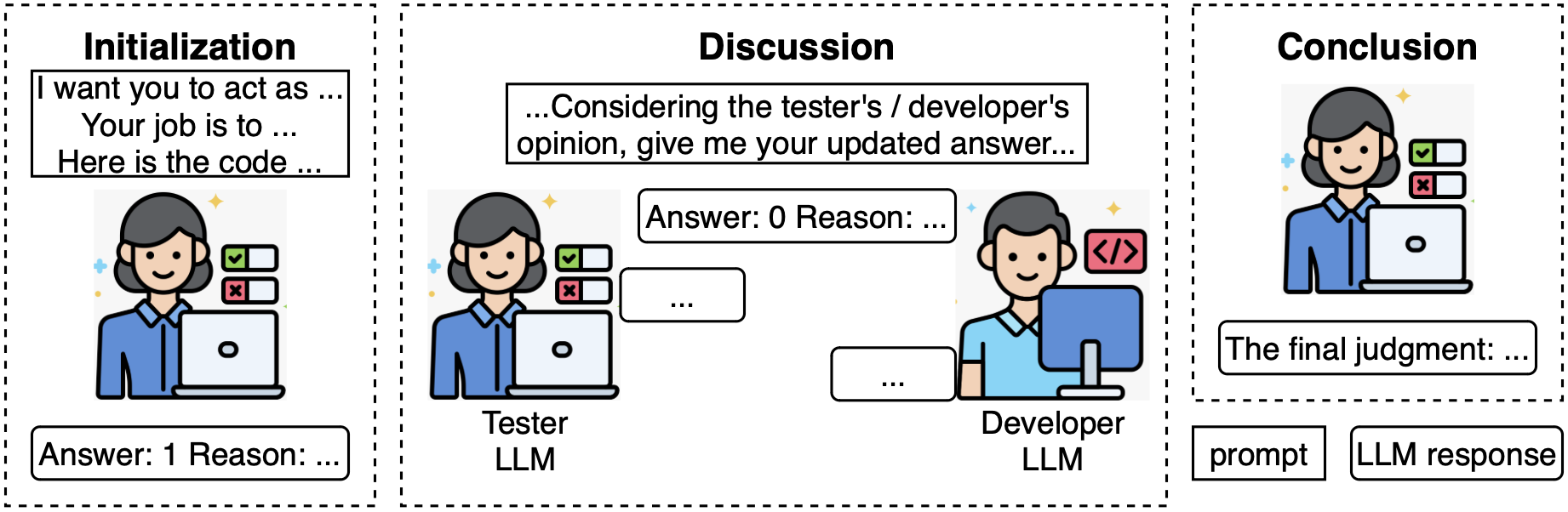}
    \caption{Overview of the multi-role approach}
    \label{fig:1}
\end{figure}

\section{Preliminary Evaluation}
The preliminary evaluation is driven by the research question as follows: In what ways does the proposed approach improve the performance of vulnerability detection?

\subsection{Experiment Settings}

\textbf{Dataset:}
The dataset used for the experiment, as referenced in \cite{Li_2022_short}, is a C/C++ dataset consisting of both vulnerable and non-vulnerable code segments across four categories: library/API function call (FC), arithmetic expression (AE), array usage (AU), and pointer usage (PU).

\textbf{Task:}
LLMs are tasked with a binary classification on whether the provided code segment contains any vulnerability that belongs to FC, AE, AU, or PU.

\textbf{Baseline:}
All LLMs used in this experiment are instances of the gpt-3.5-turbo-0125 model.
The proposed multi-role approach is compared to the result obtained from the single-role LLM approach.
Inspired by \cite{zhang2023promptenhanced_short}, the following two types of prompts are used to test these two approaches.
The basic prompt directly asks the LLM about the presence of any vulnerability in FC, AE, AU, or PU without additional context.
Chain-of-thought (CoT), i.e., step-by-step, is also used to analyze the existence of a particular vulnerability.

\textbf{Discussion constraints:}
The maximum discussion depth is set to 5, and the maximum response length is set to 120 tokens.

\textbf{Metrics:}
To assess the enhancements in vulnerability detection performance, this study employs three key metrics: the precision rate, the recall rate, and the F1 score \cite{zhang2023promptenhanced_short}.

\begin{table*}[htbp]
\centering
    \caption{Experiment Results} 
    \begin{tabular}{cccccccccccccc}\hline 
        Testing & Vulner- & \multicolumn{4}{c}{Precision rate} & \multicolumn{4}{c}{Recall rate} & \multicolumn{4}{c}{F1 score}\\
        data & ability & \multicolumn{2}{c}{single-role} & \multicolumn{2}{c}{multi-role} & \multicolumn{2}{c}{single-role} & \multicolumn{2}{c}{multi-role} & \multicolumn{2}{c}{single-role} & \multicolumn{2}{c}{multi-role} \\
        group$^1$ & category & basic & CoT & basic & CoT & basic & CoT & basic & CoT & basic & CoT & basic & CoT \\\hline
        Group1 & FC & 0.735 & 0.756 & 0.830 & \textbf{0.837} & 0.610 & 0.643 & \textbf{0.801} & 0.796 & 0.667 & 0.695 & 0.816 & \textbf{0.816}\\
        Group1 & AE & 0.750 & 0.756 & 0.837 & \textbf{0.854} & 0.618 & 0.634 & 0.779 & \textbf{0.818} & 0.677 & 0.689 & 0.807 & \textbf{0.835}\\
        Group1 & AU & 0.772 & 0.759 & \textbf{0.860} & 0.856 & 0.668 & 0.695 & 0.851 & \textbf{0.896} & 0.716 & 0.725 & 0.856 & \textbf{0.875}\\
        Group1 & PU & 0.753 & 0.769 & 0.832 & \textbf{0.847} & 0.629 & 0.654 & 0.791 & \textbf{0.875} & 0.685 & 0.707 & 0.811 & \textbf{0.861}\\\hline
        Group2 & FC & 0.568 & 0.564 & 0.637 & \textbf{0.641} & 0.640 & 0.660 & \textbf{0.752} & 0.718 & 0.602 & 0.608 & \textbf{0.690} & 0.677\\
        Group2 & AE & 0.575 & 0.553 & 0.627 & \textbf{0.647} & 0.628 & 0.640 & 0.732 & \textbf{0.750} & 0.600 & 0.593 & 0.675 & \textbf{0.694}\\
        Group2 & AU & 0.549 & 0.579 & \textbf{0.651} & 0.639 & 0.652 & 0.696 & 0.750 & \textbf{0.778} & 0.596 & 0.632 & 0.697 & \textbf{0.702}\\
        Group2 & PU & 0.550 & 0.536 & 0.619 & \textbf{0.632} & 0.674 & 0.698 & 0.778 & \textbf{0.782} & 0.606 & 0.606 & 0.690 & \textbf{0.699}\\\hline
        Group3 & FC & 0.196 & 0.204 & 0.225 & \textbf{0.232} & 0.650 & 0.695 & 0.730 & \textbf{0.765} & 0.302 & 0.315 & 0.344 & \textbf{0.356}\\
        Group3 & AE & 0.197 & 0.198 & \textbf{0.229} & 0.229 & 0.660 & 0.680 & 0.755 & \textbf{0.765} & 0.304 & 0.306 & 0.352 & \textbf{0.352}\\
        Group3 & AU & 0.218 & 0.211 & 0.238 & \textbf{0.246} & 0.735 & 0.710 & 0.805 & \textbf{0.830} & 0.336 & 0.325 & 0.368 & \textbf{0.380}\\
        Group3 & PU & 0.199 & 0.197 & 0.224 & \textbf{0.237} & 0.690 & 0.705 & 0.755 & \textbf{0.790} & 0.308 & 0.309 & 0.345 & \textbf{0.364}\\\hline
        \multicolumn{14}{l}{1 Group1 consists 800 vulnerable and 200 non-vulnerable code segments; Group2: 500 and 500; Group3: 200 and 800.}\\
    \end{tabular}
    \label{table:1}
\end{table*}

\subsection{Results and Discussions}
Table \ref{table:1} summarizes the experiment results, highlighting the improvements through the proposed approach.
On average, there is a 13.48\% increase in the precision rate, an 18.25\% increase in the recall rate, and a 16.13\% increase in the F1 score.
In terms of computation costs, due to the need for conversation between different roles, it requires a 484\% increase in the number of tokens consumed.

An illustrative discussion record demonstrating how the developer contributed to enhancing the tester's detection accuracy is as follows.
Initially, the tester identified a code segment as vulnerable due to the lack of validation or sanitization during the concatenation of two buffers.
However, as the developer pointed out that the wcsncat function and the proper size argument are used to prevent a buffer overflow, the tester finally made the correct judgment that this code segment is non-vulnerable.
Given the limitations on paper length, please refer to the complete discussion record at \url{github.com/rockmao45/LLMVulnDetection}.

The experiment results indicate a more obvious improvement, especially in the recall rate and F1 score when the proportion of vulnerable data in the testing dataset is higher.
This is likely because the rounds of discussions allows LLMs to explore a broader spectrum of potential vulnerabilities.

\section{Conclusion and Future Work}
This paper proposes a novel approach for improving vulnerability detection with LLMs, where LLMs act as different roles in a real-life code review team, discussing the existence and classification of vulnerabilities towards a consensus.
Preliminary evaluations have shown notable improvements in precision rates, recall rates and F1 scores.
Future work should enhance this approach by integrating in-context learning to guide LLMs to better collaborate in their discussions.

\bibliographystyle{IEEEtran}
\bibliography{main}

\balance

\end{document}